\documentclass[prl,twocolumn,superscriptaddress,showpacs]{revtex4}

\usepackage{amssymb}

\usepackage{graphicx}
\usepackage{dcolumn}
\usepackage{bm}
\usepackage{color}
\begin{document}
\title{Self-Assembly of Glycine on Cu (001): the Tales of Polarity and Temperature}
\date{\today}

\author{Jing~Xu}
\affiliation{Department of Physics, Renmin University of China,
Beijing 100872, China}

\author{Zheshuai~Lin}
\affiliation{Technical Institute of Physics and Chemistry, Chinese
Academy of Sciences, Beijing 100190, China}

\author{Sheng~Meng}
\affiliation{Beijing National Laboratory for Condensed Matter
Physics, Institute of Physics, Chinese Academy of Sciences, P.O.
Box 603, Beijing 100190, China}

\author{Lifang~Xu}
\email{lfxu@iphy.aphy.ac.cn.}
\affiliation{Beijing National
Laboratory for Condensed Matter Physics, Institute of Physics,
Chinese Academy of Sciences, P.O. Box 603, Beijing 100190, China}

\author{Enge~Wang}
\affiliation{School of Physics, Peking University, Beijing 100871,
China}

\begin{abstract}

Glycine on Cu(001) is used as an example to illustrate the
critical role of molecular polarity and finite temperature effect
in self-assembly of biomolecules at a metal surface. A unified
picture for glycine self-assembly on Cu(001) is derived based on
full polarity compensation considerations, implemented as a
generic rule. Temperature plays a non-trivial role: the
ground-state structure at 0~K is absent at room temperature, where
intermolecular hydrogen bonding overweighs competing
molecule-substrate interactions. The unique \emph{p}(2$\times$4)
structure from the rule is proved as the most stable one by
\emph{ab initio} molecular dynamics at room temperature, and its
STM images and anisotropic free-electron-like dispersion are in
excellent agreement with experiments. Moreover, the rich
self-assembling patterns including the heterochiral and homochiral
phases, and their interrelationships are entirely governed by the
same mechanism.

\end{abstract}
\pacs{81.16.Dn, 82.30.Rs, 68.43.Bc, 68.43.Hn}

\maketitle

Self-assembly (SA) of molecules on a solid surface is particularly
interesting for both fundamental research and potential
applications in the mass fabrication of nanoscale
devices~\cite{1,2,3,4,5,6,7,8,9,10,11,12,13}. This approach is a
promising route to construct the functional systems with nanometre
dimensions by autonomous ordering of molecules on atomically
well-defined surfaces. The comprehensive understandings of the
mechanisms controlling the SA phenomena can steer the SA and
growth processes. Various molecular patterns have been observed in
scanning tunneling microscopy (STM) for organic molecules on metal
surfaces, e.g. self-assembled supermolecular clusters and chains
on Au or Ag surfaces~\cite{3,4,5,6}; chiral metal-organic
complexes on Cu(001)~\cite{14}; and a homomolecular
two-dimensional honeycomb network at a Cu(111) surface~\cite{11}.
These organic molecules are usually planar and rigid, and directly
observable with STM. For biomolecules, however, due to their
complicated structures and properties such as chirality, it is not
possible to reach an atomic resolution in STM images, which
generally appear as bright protrusions~\cite{12,13}. For example,
recent STM studies on glycine monolayer on Cu (001) displayed a
blurred profile of molecules even at a temperature as low as 5
K~\cite{15}. Therefore, it is great desirable to investigate the
basic molecular self-assembled structures beyond the STM results.

Glycine, the simplest amino acid comprising only an amino group
(as the head) and a carboxyl group (as the tail), is one of most
fundamental components for biomolecules. Among all amino acids
glycine is the only one that does not have chirality. However, by
adsorbing on Cu surfaces glycine deprotonates and changes to
glycinate ion. Thus, the amino and carboxyl groups of glycinate
are strongly positively and negatively polarized, respectively,
upon surface adsorption, exhibiting two different chiral
configurations. Plenty of structural patterns including the
\emph{p}(2$\times$4) heterochiral phase, \emph{c}(2$\times$4)
homochiral phase, and mixed phases with different domain
boundaries have been observed for glycine on Cu (001)~\cite{16}.
To explain the stability of these various phases, Kanazawa et al.
proposed a two-step mechanism, i.e., the \emph{c}(2$\times$4)
phase is stabilized by the formation of the $\langle$310$\rangle$
steps and the \emph{p}(2$\times$4) structure by couplings of the
molecular layer to Cu substrate~\cite{16}. This model ignores the
roles of internal factors such as intermolecular hydrogen bonding
(HB) between glycine molecules, which would be the intrinsic
driving force for SA. Scanning tunneling spectroscopy (STS)
further revealed anisotropic free-electron-like dispersions with
electron effective mass differing by 10-fold along the [110] and
[$\bar{1}$10] directions, respectively~\cite{15}. Although with
many tries~\cite{17,18}, building an atomistic model of glycine
monolayer structure that is thermodynamically stable and
reproducing this anisotropic electronic behavior has not been
successful. The mechanism governing the self-assembly of glycine
on Cu, which has been a long-term controversy, is still unclear.

In this work, we employed state-of-the-art first-principles
calculations and extensive molecular dynamics (MD) simulations to
investigate the SA structure of glycine on Cu(001). We discovered
that the structure satisfying a complete polarity compensation
reproduces well all known features observed in experiments
including STM images and 10-fold anisotropic free-electron-like
dispersion, and further predicted all possible homochiral and
mixed phases for glycine on Cu(001). Our results indicate that
molecular polarity and finite temperature play vital roles in
determining the microscopic mechanisms for glycine SA on a metal
substrate. The ``ground-state structure'' at 0 K is not stable at
room temperature; instead, thermal fluctuations weaken
molecule-substrate interaction and favor intermolecular hydrogen
bonds as determinative forces for glycine self-assembling on Cu
surface, which is a unique intrinsic nature of self-assembly in
this system.

Structure optimizations were performed within the framework of
Density Functional Theory (DFT) using Vienna \emph{ab initio}
simulation package (VASP)~\cite{19,20}, in which
projector-augmented wave pseudopotentials and generalized gradient
approximation~\cite{21} were chosen. A plane wave basis set was
used to expand the Kohn-Sham orbitals with a 400 eV kinetic energy
cutoff. Glycine molecules were placed on a seven layer Cu(001)
slab in a supercell with dimensions of
10.28\AA$\times$10.28\AA$\times$25\AA. The Monkhorst-Pack
scheme~\cite{22} was adopted for the Brillouin zone integration,
and we have tested that a 4$\times$4$\times$1 \emph{k}-point mesh
is sufficient to ensure a good convergence in the total energy
differences. The atoms in the top four Cu layers were allowed to
fully relax until the forces on them were all smaller than 0.03
eV/\AA. Energy convergence for the geometry optimization was
better than 0.1 meV per atom. The time step in molecular dynamics
simulations was chosen to be 0.5 \emph{f}s. An equilibration
process was first performed by slowly heating the systems from 0~K
to 310 K. Then we examined thermal oscillations in atomic
structure and free energy at 310~K in canonical ensemble using the
Nos\'{e} algorithm.

\begin{figure}
   \includegraphics[width=8cm]{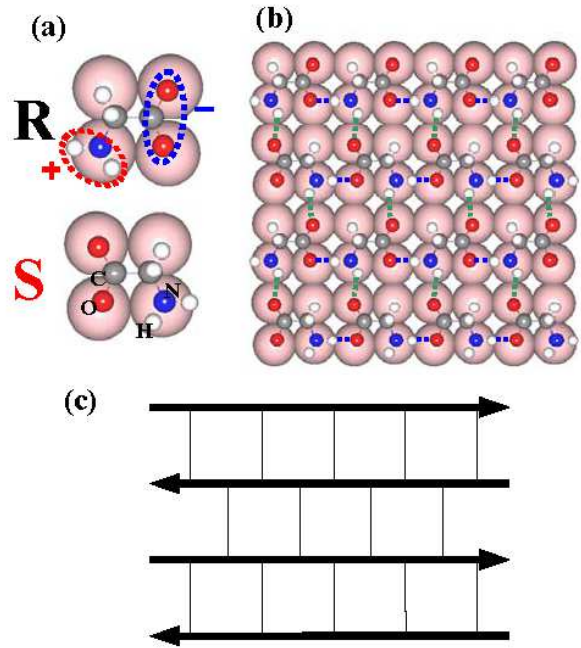}\\
   \caption{(a) Adsorbed glycinate molecule (\emph{R}) and its enantiomeric isomer (\emph{S}).
    Two positive sites in amino group and two negative sites in carboxyl group are circled by red
    and blue dashed lines, respectively. (b) The anti-parallel molecular arrangement of a
    heterochiral \emph{p}(2$\times$4) structure. Each molecule interacts with four neighbors
     via HB. (c) Schematic plot of anti-parallel HB ladder-like network. Horizontal bold lines
     represent strong HB along [110] and vertical thin lines weak HB along [$\bar{1}$10].
     The arrows indicate the HB directions.}\label{fig1}
\end{figure}

Our first-principles calculations show that a single glycinate
molecule and its enantiomeric isomer binds to the Cu (001) surface
with an energy of 2.33 eV through one nitrogen atom and two oxygen
atoms, all on top site of substrate Cu atoms, as schematically
shown in Fig.~\ref{fig1}(a). Then we examined the atomistic
geometry of glycinate SA monolayers. Based on physical
consideration and the characters of glycinate molecule, we found
that the stable \emph{p}(2$\times$4) structures have the following
features: First, due to strong attractions between the positively
(amino head) and negatively polarized ends (carboxyl tail),
glycinate molecules arrange themselves into a linear chain via
head-to-tail HB along the [110] direction; Second, these chains
form an alternating anti-parallel pattern to eliminate the net
polarity of the two-dimensional (2D) island; Third, the
neighboring chains are also connected by weaker HB along the
[$\bar{1}$10] direction to make the monolayer structure an
integral network. The above three polarity compensation reasons
emphasize that the constructing glycinate SA arrangement is
energetically favorable by maximizing intermolecular attractive
interactions while removing the net polarity of 2D network on a
surface. Under these restrictions, we \emph{exclusively} reached a
heterochiral \emph{p}(2$\times$4) structure of glycinate on
Cu(001), as shown in Fig.~\ref{fig1}(b). Glycinate molecules
naturally alternate their chirality from one chain to the next, as
required by the full polarity compensation, otherwise, less stable
and isolated double-rows would form instead of the 2D monolayers
as observed in experiment.

Geometry optimization shows that all O...H lengths in the HB along
[11] are 1.6~\AA\ with the binding energy of 210 meV (indicating a
strong HB), whereas those along the [$\bar{1}$10] direction are
2.0~\AA\ with the energy of 138 meV (weak HB)~\cite{23}. They are
denoted in blue and green dashed lines in Fig.~\ref{fig1}(b),
respectively. This anti-parallel HB network is also schematically
drawn in Fig.~\ref{fig1}(c), to indicate the arrangement of
molecular polarity. In this structure each glycinate acts as
\emph{two HB donors} and \emph{two HB acceptors}, where all its
polar sites are nicely saturated with HB~\cite{24}. This is
consistent with the general understanding of H bonding
interactions: namely, the selectively, linearity, and saturation
of H bonds.

\begin{figure}
   \includegraphics[width=8cm]{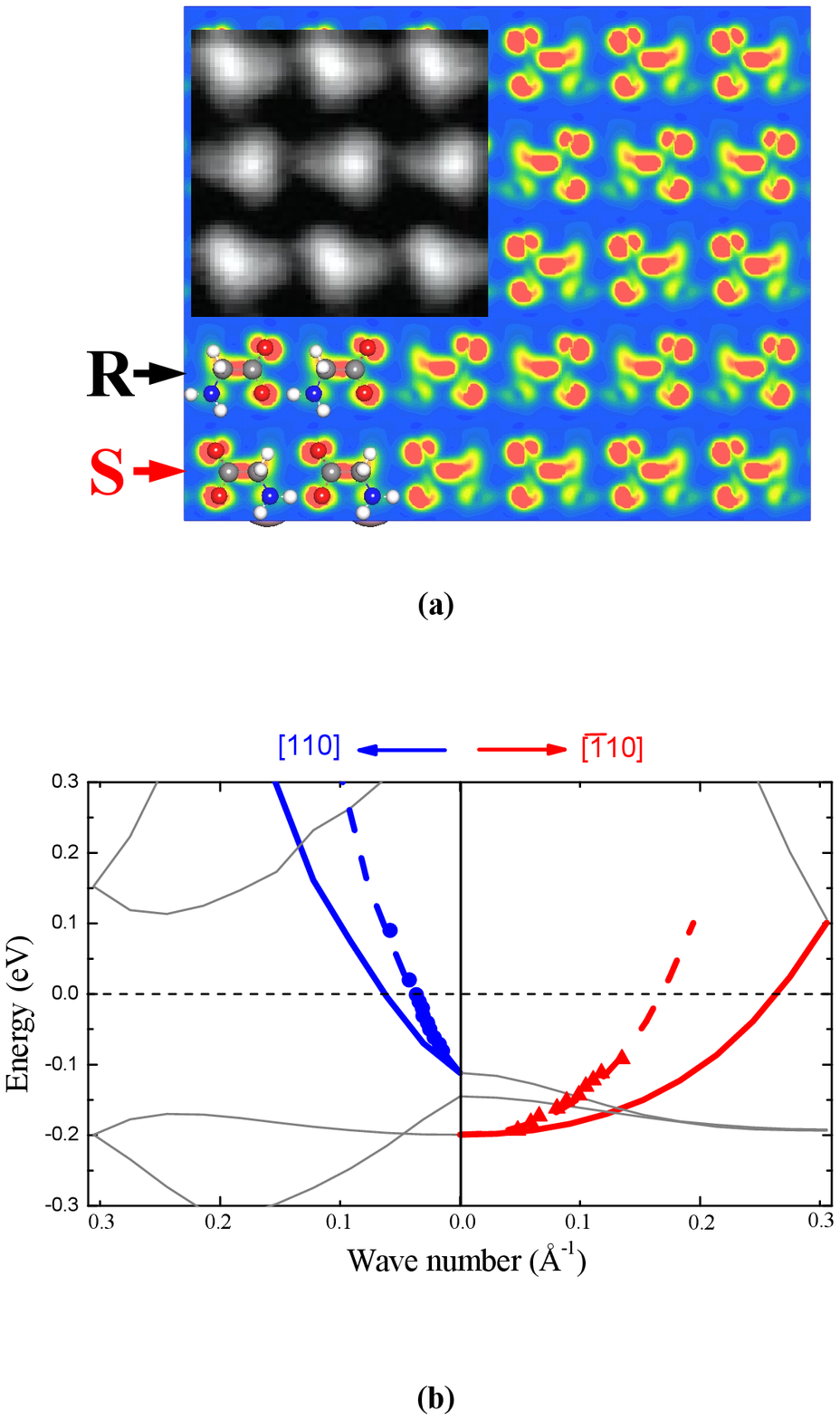}\\
   \caption{(a) The calculated STM image for the \emph{p}(2$\times$4) structure,
   the insert is experimental result from Ref. [15] for comparison. (b) Energy
    dispersion relations for the [110] direction in blue and [$\bar{1}$10] in red,
    respectively. Solid lines represent our calculated results and dashed lines
    are rescaled for comparison with experimental results from Ref. [15]
    (blue dots for [110] and red triangles for [$\bar{1}$10] ).
    The Fermi energy is indicated by horizontal dashed line.}\label{fig2}
\end{figure}

Any structures proposed for glycine SA on Cu(001) would have to
withstand stringent tests to reproduce properties measured
experimentally; our structure model based on polarity
compensation, shown in Fig.~\ref{fig1}, satisfies these tests very
well. To confirm it is indeed the one observed in experiment, we
compare STM images, electronic band structure, structural phases,
and energetics from both theory and experiment. Fig.~\ref{fig2}(a)
shows the simulated STM image for the \emph{p}(2$\times$4) phase,
together with the measured image~\cite{15} in the inset for
comparison. The STM image is calculated by integrating local
states with energy $<$100 meV below the Fermi level within
Tersoff-Hamann approximation~\cite{25}. It displays a pattern of
triangular protrusions: triangles slightly tilted to the left are
glycinate in \emph{S} conformation, while those tilted to the
right are in \emph{R} conformation, forming an array of
alternating anti-parallel rows. In experiment a pattern of blurred
triangles with two orientations was observed, forming
anti-parallel alternating rows along [110], in a very similar
manner to simulated images. Within each row the molecules are
close to each other, indicating a stronger interaction within the
rows; while there is a larger separation between the rows,
suggesting weaker row-to-row interactions. These features are well
reproduced and explained by our model.

More stringent test comes from the 10-fold difference in electron
effective mass $m^{*}_{e}$ for the free-electron-like dispersions
along the [110] and [$\bar{1}$10] directions observed in
STS~\cite{15}. Again this strong anisotropic behavior is obtained
for our proposed \emph{p}(2$\times$4) structure. We calculate the
surface band structure of glycinate on a single Cu(001) layer, in
order to remove the noisy background of bulk substrate. This
approximation does not affect the anisotropic property because the
removed Cu atoms are symmetric to the [110] and [$\bar{1}$10]
directions. The results are shown in Fig.~\ref{fig2}(b). There is
only one band across Fermi surface for both [110] and
[$\bar{1}$10] directions. The solid lines represent the calculated
results, and the dashed lines are rescaled by multiplying a
constant 2.46 for direct comparison with experimental data. Dashed
curves perfectly match the experimental results, showing parabolic
free-electron-like dispersions, as well as the 10-fold difference
in the effective mass with respect to the [110] and [$\bar{1}$10]
directions. The scaling constant is attributed to an enlarged
effective mass $m^{*}_{e}$ by using a single layer Cu in our
approximation. Meanwhile, the Fermi energy is also shifted to
include contributions from Cu layers underneath. Clearly, the
anisotropy originates from different strengths of interactions
between neighboring molecules in two directions, i.e., strong HB
along the direction [110] and weak HB along [$\bar{1}$10], shown
schematically in Fig.~\ref{fig1}(c). Note that the weaker HBs are
misaligned along  [$\bar{1}$10] which would significantly
decreases the conductivity in this direction as well. Thus,
different HB strengths together with their arrangements result in
the 10-fold difference in the effective mass.

To interpret their STM observations, Kanazawa et al. adopted a
structural model consisting of only head-to-tail alignments in the
same direction, which is prevailing in literature for glycine SA
on Cu(001)~\cite{12,15,16,26}. For convenience of presentation, we
name this structure the head-to-tail (HT) model. This model,
however, does not consider the influence of polarity to side
neighbors, not mention to the HB network. Thus, neither STM images
nor free-electron-like dispersion in experiments were reproduced
by the HT model. Indeed, first-principles calculation shows that
the HT structure is 76~meV higher per molecule in energy than our
structure for \emph{free-standing} monolayers, indicating that our
model is intrinsically more stable for SA than the HT model. In
our \emph{p}(2$\times$4) structure glycinate molecules are
saturated with H bonds and all H bonds conform the ideality of H
bonding interactions such as selectively and linearity, thus its
free-standing structure is more stable than the unsaturated HT
structure. The apparent -CH...O hydrogen bonds with the length of
2.3 \AA\ in the HT structure cannot be regarded as an ideal HB.

However, a complexity arises from the subtle balance between the
intralayer molecule-molecule interactions and the
molecule-substrate binding, which is a central theme in all SA
processes. At 0~K, first principles calculations show that the HT
model is lower in total energy by 140 meV per glycinate than the
adsorption of our structure based on polarity compensation. That
is, the HT model is the ``ground state structure'' for glycinate
layer on Cu(001) at 0~K. Then why it is not observed in
experiments, which shows very different STM/STS features as to
those of the HT structure?

\begin{figure}
   \includegraphics[width=8cm]{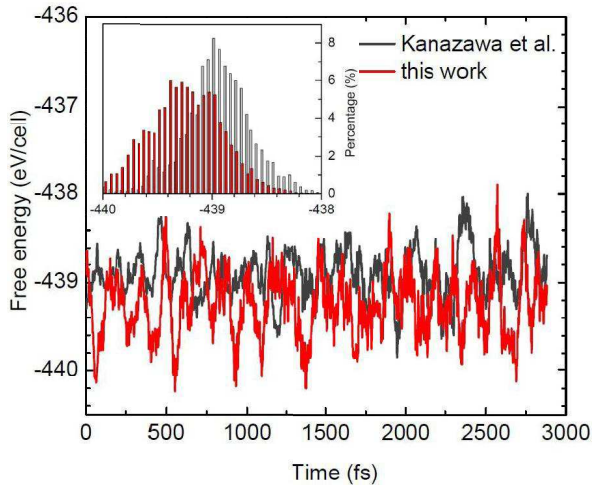}\\
   \caption{Free energy fluctuations as a function of time in molecular
    dynamics simulations at 310 K for our structure and that adopted by
    Kanazawa et al. (Refs. [15,16]). The inset shows the distribution of
    free energies for the two cases.}\label{fig3}
\end{figure}

To solve this controversy, we performed \emph{ab initio} MD
simulations for both HT and our structure model at room
temperature (RT)$\sim$310~K. Both configurations are stable in
trajectories lasting for 5 ps. However, we found the ``ground
state structure'' at 0~K presents a higher free energy at 310~K,
which is now 75~meV per molecule higher in energy than our model,
shown in Fig.~\ref{fig3}. The difference in free energy is
coincidental to that for free-standing layers, 76~meV, suggesting
that the SA structure at RT is dominated by intermolecular
H-bonds, rather than the molecule-substrate bonds as in the case
at 0~K. At absolute zero the \emph{unsaturated} HT model has
stronger interaction with the substrate compared to our
\emph{saturated} structure. Thermal fluctuations weaken
glycinate-Cu bonds, and free glycinates adjust themselves to
maximize HB interactions while maintaining glycine-Cu interaction
in a similar strength.

In fact, besides the molecule-molecule and molecule-surface
interactions, the temperature plays as the third important
parameter in molecular self-assembly at surfaces~\cite{4}. During
the SA process, molecules interact with each other sufficiently
with the help of thermal energy. In experimental conditions
(temperature $\sim$ 370~K\cite{15} and $\sim$ 430~K\cite{12}),
glycinate molecules move around on Cu surface in translational and
rotational modes, and interact with each other by HB. The HB would
break and re-bond, providing sufficient opportunities for
molecules to search for stronger intermolecular interactions. This
process, together with strong polarity compensation of glycinate,
promotes formation of anti-parallel HB network on Cu(001). Once
formed, the structure is trapped as a metastable phase even at
very low temperatures. The fact that the structure we proposed is
observed at a temperature as low as 5 K in experiments, implies
that transition from our structure to the ``ground-state'' HT
structure could be blocked by a large energy barrier. We emphasize
that the SA structure at a finite temperature is determined by the
free energy, rather than the total energy at 0~K. The
molecule-surface interaction decides the adsorbed glycinate
molecular configuration in the tridentate fashion, but the
intermolecular interaction plays a dominant role for the glycinate
SA arrangements on Cu(001). Therefore, it is demonstrated that at
finite temperature the SA on surfaces are mainly determined by the
relatively weak HB rather than the stronger interfacial
interaction.

Based on above investigations, we summarize a general rule for the
construction of glycinate SA structures on Cu surface. The rule
has three elementary components: (i) Because of the strong
polarity, the positive site of a glycinate interacts with the
negative site of another via hydrogen bond to form a head-to-tail
chain. (ii) The polarity of glycinate also affects its side
neighbors, i.e., the positive sites of neighbors will approach to
the negative site of the glycinate, and vise versa. Thus, (i) and
(ii) together result in the \emph{local} molecular configurations
where the positive site of each glycinate is surrounded by the
three neighbors' negative sites, and the negative site surrounded
by the three positive sites. (iii) Each glycinate molecule
contributes its all active sites, \emph{two positive sites} in
amino group and \emph{two negative} sites in carboxyl group, to
interact with its neighbors via HB. According to the rule (iii),
glycinate molecule will automatically adjust its chirality, an
important feature associated with SA, to self-organize HB network
structures. These three principles for constructing glycinate SA
are collectively called the polarity compensation rule here.

Besides the dominant heterochiral \emph{p}(2$\times$4) phases
widely observed in experiments~\cite{12, 15}, homochiral phases
were also found by several groups for glycinate SA on
Cu(001)~\cite{12, 16}. The presence of both heterochiral and
homochiral phases induces a variety of SA patterns in STM
images~\cite{16}, e.g. a homochiral array at the boundary between
heterochiral \emph{p}(2$\times$4) phases. To reach a unified
understanding on these rich SA patterns, we invoke the polarity
compensation rule to unravel the underlying mechanism for the
formation of different chiral phases.

\begin{figure}
   \includegraphics[width=8.9cm]{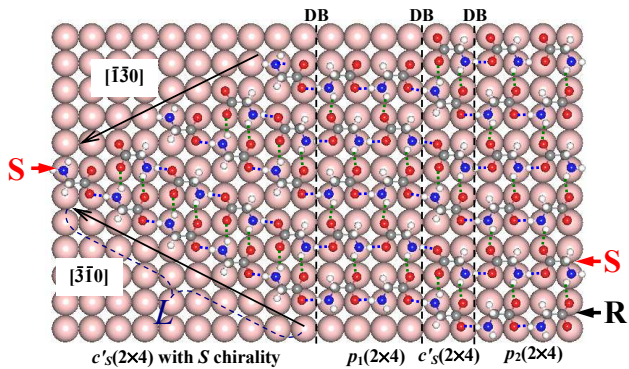}\\
   \caption{Molecular arrangements for comparing with STM images in Ref. [16],
   where exhibit various structures including the heterochiral \emph{p}(2$\times$4)
   and homochiral \emph{c}$_{S}^\prime$(2$\times$4) domains and their relationship.
   Dashed lines indicate domain boundaries.}
   \label{fig4}
\end{figure}

We first look at the boundary between two \emph{p}(2$\times$4)
domains, \emph{p}$_1$(2$\times$4) and \emph{p}$_2$(2$\times$4),
where \emph{p}$_2$ is shifted by one lattice of Cu surface along
[$\bar{1}$10] with respect to $p_1$, shown in Fig.~\ref{fig4}. At
the boundary, the glycinate with \emph{R} chirality has to turn
its head into \emph{S} chirality in order to form a strong HB
across the boundary and a weak HB with the side neighbor. Thus all
molecules at the leftmost column of \emph{p}$_2$(2$\times$4)
become \emph{S} chirality, denoted as
\emph{p}$_{S}^\prime$(2$\times$4) in Fig.~\ref{fig4}. This
homochiral glycinates array must exist between the shifted
heterochiral \emph{p}$_1$ and \emph{p}$_2$ domains, offering full
HBs to stabilize the boundary and lowering the energy by $\sim$100
meV per molecule. This simple mechanism provides a natural
explanation for the well observed homochiral phase between
\emph{p}(2$\times$4) domains in experiment, indicated by white
arrows in Fig.~2(e) of Ref. [16].

\begin{figure*}
   \centering
   \includegraphics[width=12.5cm]{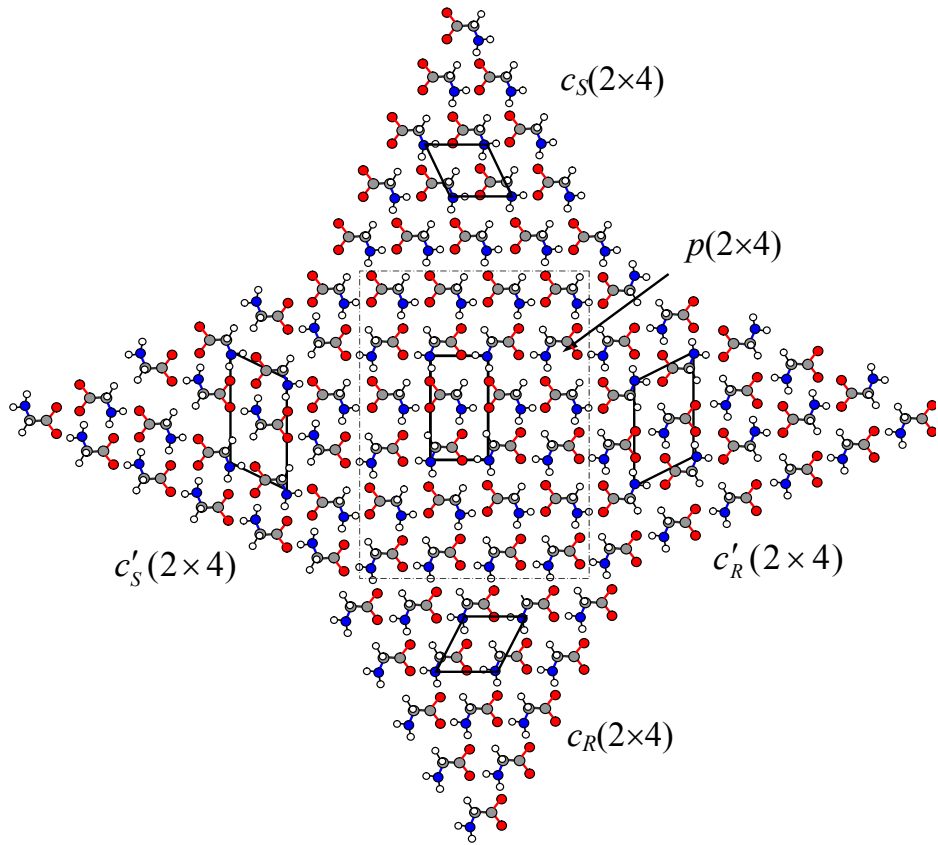}\\
   \caption{Panorama of the SA monolayer of glycinate on Cu (001).
   There are four homochiral phases around the heterochiral phase \emph{p}(2$\times$4).
   Domain boundaries are represented by the dashed lines, and the unit
   cell for each phase is denoted by solid lines. For the sake of clarity,
   the Cu substrate is not shown.}
   \label{fig5}
\end{figure*}

The above observation demonstrates how glycinate changes its
chirality to interact with its neighbors according to the rule
when the molecule shifts along [$\bar{1}$10] by a Cu position.
Shifting continuously in this way, all glycinates with \emph{R}
chirality would change to \emph{S} chirality in order to bond
together, so a complete homochiral phase of \emph{S} chirality is
formed. More importantly, this homochiral phase naturally develops
a step along the [$\bar{3}$$\bar{1}$0] direction, on which all
molecules are saturated with HB and the step becomes inactive,
shown in the left corner of Fig.~\ref{fig4} and Fig.~\ref{fig5}.
This triangular \emph{p}$_{S}^\prime$(2$\times$4) phase is indeed
observed by STM, denoted as \emph{c}(2$\times$4) in Fig.~2(e) of
Ref.~[16].

Since glycinate is enantiomeric, it is expected that an
alternative homochiral \emph{c}$_{R}^\prime$(2$\times$4) phase
with all \emph{R} chirality also exists. Indeed this homochiral
\emph{R}-phase is formed in the same way at the right side of the
\emph{p}(2$\times$4) pattern, illustrated in Fig.~\ref{fig5}.
Similarly, this phase could also be formed by shifting the
glycinate molecules along [1$\bar{1}$0] at the left side of the
\emph{p}(2$\times$4) pattern by a Cu position, while the
\emph{c}$_{S}^\prime$(2$\times$4) phase at the right side. Both
\emph{c}$_{S}^\prime$ and \emph{c}$_{R}^\prime$ phases are
composed of hydrogen-bonded twin chains. Besides the
aforementioned homochiral phases, by shifting the glycinate a Cu
position along [110] sequently, other two homochiral phases above
and below the \emph{p}(2$\times$4) pattern can also be constructed
based on our rule, named as \emph{c}$_{S}$(2$\times$4) and
\emph{c}$_{R}$(2$\times$4) in the upper and lower part of
Fig.~\ref{fig5}, respectively. The chirality of the outermost
glycinate in the \emph{p}(2$\times$4) phase decides the chirality
of the \emph{c} phase. Since each adsorbed glycinate molecule
occupies a (2$\times$2) Cu surface cell, every possibility for the
\emph{c}' and \emph{c} phases by shifting a Cu position along each
direction has been considered. It should be noted that all above
homochiral phases are different from the c(2$\times$4) phase model
adopted by Kanazawa et al.\cite{16}.

We have exhausted all possible SA structures of glycinate on
Cu(001): a unique heterochiral \emph{p}(2$\times$4) phase and four
homochiral phases based on our rule. In particular, the chirality
of each phase is automatically determined by the rule (iii). The
interrelationships between various phases are displayed in
Fig.~\ref{fig5} where the heterochiral \emph{p}(2$\times$4) phase
is surrounded by different homochiral phase in each direction. The
reason that we put them together is to emphasize their
\emph{seamless} connections where the boundaries between the
hetero- and homo-chiral phases also obey our rule. No matter where
a glycinate molecule locates in single phase or boundary, it fully
interacts with the neighbors by two positive sites in amino group
and two negative sites in carboxyl group via HB without any
exception. To our knowledge, glycine on Cu(001) exhibits the
richest 2D SA structures on surface, in which all SA patterns and
their interrelationships are naturally explained under a single
mechanism. Since \emph{p}(2$\times$4) and
\emph{c}$_{S}^\prime$(2$\times$4) structures have already been
observed in STM experiment\cite{16}, searching for other predicted
phases experimentally remains a challenge to prove our theory.

To conclude, we have clarified that the self-assembly structure is
determined by the free energy at finite temperature, and revealed
the mechanism for self-assembly of glycine on Cu(001) based on
full polarity compensation. The ideality of H bonding environment
reaches a unique antiparallel head-to-tail network of glycinates
in \emph{p}(2$\times$4) phase. We elucidate that thermal
fluctuations at finite temperatures weaken the molecule-substrate
interaction and favor the intermolecular interactions, which plays
a dominant role to orchestrate glycine self-assembly. The
long-term controversy about competing phases, i.e., homechiral
phase and heterochiral phase, as well as their relationships has
been uniformly resolved in the same framework. The present work
represents a distinct example that self assembly of molecules is
driven by intermolecular interactions themselves, rather than
being fixed by interacting with templates, even though the
molecule-template interaction is stronger at 0~K. Although
demonstrated for a small molecule on a simple surface, the SA
mechanism and the rule revealed here are expected to have much
wider implications in understanding self-assembly of polar
molecules with other templates and of larger biomolecules in
biologically relevant environments~\cite{4,5}.


\begin{thebibliography}{200}
\bibitem{1}
J. V. Barth, G. Costantini, and K. Kern, Nature (London) {\bf
437}, 671 (2005).
\bibitem{2}
F. H\"{o}\"{o}k et al., ACS Nano 2008, {\bf 2}, 2428-2346.
\bibitem{3}
M. B\"{o}hringer et al., Phys. Rev. Lett. {\bf 83}, 324 (1999).
\bibitem{4}
J. V. Barth et al., Angew. Chem. Int. Ed. {\bf 39}, 1230 (2000).
\bibitem{5}
J. Weckesser et al., Phys. Rev. Lett. {\bf 87}, 096101 (2001).
\bibitem{6}
A. K\"{u}hnle et al., Phys. Rev. Lett. {\bf 93}, 086101 (2004).
\bibitem{7}
T. Yokoyama et al., Nature {\bf 413}, 619-621 (2001).
\bibitem{8}
S. Stepanow et al., Nat. Mat. {\bf 3}, 229-233 (2004).
\bibitem{9}
J. A. Theobald et al., Nature  {\bf 424}, 1029-1031 (2003).
\bibitem{10}
R. Madueno et al., Nature {\bf 454}, 618-621 (2008).
\bibitem{11}
G. Pawin et al., Science {\bf 313}, 961-962 (2006).
\bibitem{12}
(a) X. Zhao, et al., Surf. Sci. {\bf 424}, L347-L351 (1999); (b)
X. Zhao, et al., Mater. Sci. Eng. C {\bf 16}, 41-50 (2001).
\bibitem{13}
V. Efstathiou and D. P. Woodruff, Surf. Sci. {\bf 531}, 304
(2003).
\bibitem{14}
P. Messina et al., J. Am. Chem. Soc. {\bf 124}, 14000-14001
(2002).
\bibitem{15}
K. Kanazawa et al., J. Am. Chem. Soc. {\bf 129}, 740-741(2007).
\bibitem{16}
K. Kanazawa et al., Phys. Rev. Lett. {\bf 99}, 216102 (2007).
\bibitem{17}
M. S. Dyer and M. Persson, J. Phys.: Condens. Matter {\bf 20},
312002 (2008)
\bibitem{18}
Z. X. Hu, W. Ji and H. Guo, Phys. Rev. B {\bf 84}, 085414 (2011).
\bibitem{19}
G. Kresse and J. Hafner, Phys. Rev. B {\bf 47}, 558 (1993).
\bibitem{20}
G. Kresse and J. Furthm\"{u}ller, Phys. Rev. B {\bf 54}, 11169
(1996)
\bibitem{21}
Y. Wang and J. P. Perdew, Phys. Rev. B {\bf 44}, 13298 (1991).
\bibitem{22}
H. J. Monkhorst and J. D. Pack, Phys. Rev. B {\bf 13}, 5188
(1976).
\bibitem{23}
S. Meng et al., Phys. Rev. Lett. {\bf 89}, 176104 (2002).
\bibitem{24}
J. J. Yang et al., Phys. Rev. Lett. {\bf 92}, 146102 (2004).
\bibitem{25}
J. Tersoff and D. R. Hamann, Phys. Rev. B {\bf 31,} 805 (1985).
\bibitem{26}
K. Mae and Y. Morikawa, Surf. Sci. {\bf 553}, L63 (2004).


\end{thebibliography}
\end{document}